\documentclass[aps,prl,twocolumn,nofootinbib,showpacs,floatfix,superscriptaddress,preprintnumbers]{revtex4}
\usepackage{amsmath}
\usepackage{amssymb}
\usepackage{epsfig}
\usepackage{graphicx}
\usepackage{pstricks}
\usepackage{pst-coil}
\DeclareMathAlphabet{\mathsc}{OT1}{cmr}{m}{sc}
\newcommand {\ignore}[1]{}

\renewcommand{\baselinestretch}{1.1}

\def\10{$SO(10)$}
\def\21{SU(2) $\otimes$ U(1) }

\def\422{$SU(4) \otimes SU(2) \otimes SU(2)$}
\def\321{SU(3) $\otimes$ SU(2) $\otimes$ U(1)}
\def\gsim{\raise0.3ex\hbox{$\;>$\kern-0.75em\raise-1.1ex\hbox{$\sim\;$}}}
\def\lsim{\raise0.3ex\hbox{$\;<$\kern-0.75em\raise-1.1ex\hbox{$\sim\;$}}}

\def\lsim{\raise0.3ex\hbox{$\;<$\kern-0.75em\raise-1.1ex\hbox{$\sim\;$}}}
\def\gsim{\raise0.3ex\hbox{$\;>$\kern-0.75em\raise-1.1ex\hbox{$\sim\;$}}}
\def\vev#1{\left\langle #1\right\rangle}
\def \znbb {0\nu\beta\beta}

\newcommand{\AddrAHEP}{%
  AHEP Group, Institut de F\'{\i}sica Corpuscular --
  C.S.I.C./Universitat de Val{\`e}ncia \\
  Edificio Institutos de Paterna, Apt 22085, E--46071 Valencia, Spain}

\renewcommand{\baselinestretch}{1.25}
 \newcommand{\ba}{\begin{array}}
\newcommand{\ea}{\end{array}}
\relax

\def\321{$SU(3)\times SU(2)\times U(1)$}

\newcommand{\Sol}  {\textrm{sol}}

\newcommand{\Atm}  {\textrm{atm}}

\begin{document}
\preprint{IFIC/08-50}
\renewcommand{\Huge}{\Large}
\renewcommand{\LARGE}{\Large}
\renewcommand{\Large}{\large}
\def \znbb {$0\nu\beta\beta$ }
\def \nbb {$\beta\beta_{0\nu}$ }
\title{Modelling tri-bimaximal neutrino mixing }
\author{M.~Hirsch} \email{mahirsch@ific.uv.es} 
\author{S.~Morisi} \email{ morisi@ific.uv.es}
\author{J.~W.~F.~Valle} \email{valle@ific.uv.es}
\affiliation{\AddrAHEP}

\date{\today}

\begin{abstract}

  We model tri-bimaximal lepton mixing from first principles in a way
  that avoids the problem of the vacuum alignment characteristic of
  such models. This is achieved by using a softly broken $A_4$
  symmetry realized with an isotriplet fermion, also triplet under
  $A_4$.  No scalar $A_4$-triplet is introduced. This represents one
  possible realization of general schemes characterized by the minimal
  set of either three or five physical parameters. In the three
  parameter versions $m_{ee}$ vanishes, while in the five parameter
  schemes the absolute scale of neutrino mass, although not predicted,
  is related to the two Majorana phases. The model realization we
  discuss is potentially testable at the LHC through the peculiar
  leptonic decay patterns of the fermionic and scalar triplets.

\end{abstract}


\maketitle

So far the historical discovery of neutrino oscillations constitutes
the only indication of physics beyond the standard model.  It is
especially intriguing that the pattern of neutrino mixing
angles~\cite{Schwetz:2008er} implied by current neutrino data is in
sharp contrast with the structure of the Cabibbo-Kobayashi-Maskawa
quark mixing matrix~\cite{Kobayashi:1973fv}.
Current data do not yet fully determine the absolute scale of neutrino
masses, nor do they shed any light on the important issue of leptonic
CP violation~\cite{Nunokawa:2007qh,Bandyopadhyay:2007kx}.

Lacking a basic theory of flavor and the origin of mass it is useful
to have theoretical models restricting the pattern of fermion masses
and mixings and providing guidance for future experimental searches.
A successful phenomenological ansatz for leptons has been proposed by
Harrison, Perkins and Scott (HPS) and is given
by~\cite{Harrison:2002er}
\begin{equation}
\label{eq:HPS}
U_{\textrm{HPS}} = 
\begin{pmatrix}
\sqrt{2/3} & 1/\sqrt{3} & 0\\
-1/\sqrt{6} & 1/\sqrt{3} & -1/\sqrt{2}\\
-1/\sqrt{6} & 1/\sqrt{3} & 1/\sqrt{2}
\end{pmatrix}
\end{equation}
which corresponds to 
$$\tan^2\theta_{\Atm}=1,~~~\sin^2\theta_{\textrm{Chooz}}=0,~~~\tan^2\theta_{\Sol}=0.5,$$
providing a good first approximation to the values indicated by
current neutrino oscillation data.

In this note we discuss the general form of the effective neutrino
mass operator which yields tri-bimaximal lepton mixing $U_{\textrm{HPS}}$. In general it
is characterized by 5 physical parameters. These determine the two
neutrino mass square differences measured in current neutrino
oscillation experiments. The third neutrino mass parameter, namely the
absolute scale of neutrino masses, though not strictly predicted,
correlates with the remaining two free parameters, namely two Majorana
phases. Hence these would be indirectly tested by probing the absolute
scale of neutrino mass in beta decay endpoint studies, neutrinoless
double beta decay (\znbb)
searches~\cite{Avignone:2007fu,Hirsch:2006tt}, or cosmology
\cite{Lesgourgues:2006nd}.

We provide a possible gauge-theoretic realization based on the
non-abelian discrete $A_4$ flavour symmetry. 
All $A_4$ models predicting $U_{\textrm{HPS}}$ lepton mixing, need at least
two $A_4$-triplets whose vacuum expectation values (vevs) have
different $A_4$-alignments. Different papers have emphasized this
problem \cite{Ma:2008ym,Grimus:2008tt,Zee:2005ut,Ma:2004zv}.  In order to account for such
alignments, extra-dimensions~\cite{Altarelli:2005yp,Kobayashi:2008ih}
and/or supersymmetry~\cite{Babu:2005se,Altarelli:2005yx} have been
invoked.
Here we show how $U_{\textrm{HPS}}$ mixing can be obtained with just 
$A_4$ Higgs singlets in the framework of a softly broken $A_4$ theory. The
model is similar in spirit, but inequivalent to the one suggested in
Ref.~\cite{Ma:2008ym}.


We start by noting that if $M_\nu$ is
\begin{equation}\label{mnu}
M_\nu =\left(
\begin{array}{ccc}
x(1+\alpha)&y(1+\alpha)&y(1+\alpha)\\
y(1+\alpha)&x+\alpha y&y +\alpha x\\
y(1+\alpha)&y +\alpha x &x+\alpha y
\end{array}
\right)
\end{equation}
in the basis where charged leptons are diagonal, the lepton mixing
matrix is $U_{\textrm{HPS}}$ independently of the values of $x,\,y$
and $\alpha$, since it is $\mu\leftrightarrow \tau$ exchange invariant
and $M_{\nu_{11}}=M_{\nu_{22}}+M_{\nu_{23}}-M_{\nu_{13}}$.  Here $x,\,
y$ and $\alpha$ are three complex free parameters.  In such
``weak-basis'' the physical parameters can be chosen as their
\textit{moduli} $|x|, \,|y|,\, |\alpha|$, the phase of $\alpha$, which
we denote with $\delta_1$ and the relative phase between $x$ and $y$,
denoted as $\delta_2$.  The eigenvalues of $M_\nu$ are
\begin{eqnarray}
m_1&=&(1+\alpha)(x-y),\nonumber\\
m_2&=&(1+\alpha)(x+2y),\nonumber\\
m_3&=&(1-\alpha)(x-y).
\end{eqnarray}
Note that, the vanishing of $\theta_{13}$ removes the possibility of
Dirac-type leptonic CP violation. Hence the two phases $\delta_1$ and
$\delta_2$ are directly (but not trivially) related to the two
physical Majorana phases of the lepton mixing
matrix~\cite{schechter:1980gr}.
On the other hand we can rewrite the \textit{moduli} $|x|, \,|y|,\,
|\alpha|$ as the three neutrino mass parameters, which may be chosen
as the two mass square splittings that enter the oscillations,
\begin{equation}\label{syst}
\begin{array}{l}
 |m_3|^2-|m_1|^2 = \Delta m^2_{atm}=\\
\quad=-4|\alpha| \cos\delta_1(|x|^2+|y|^2-2|x||y|\cos\delta_2),\\
\\
|m_2|^2-|m_1|^2=\Delta m^2_{sol}=\\
\quad=3(1+2|\alpha| \cos \delta_1 + |\alpha|^2)(2|x||y|\cos\delta_2+|y|^2),
\end{array}
\end{equation}
plus the absolute
mass scale that governs neutrinoless double beta decay
\begin{equation}\label{mee}
\begin{array}{l}
|m_{ee}|^2=|x|^2(1+2 |\alpha| \cos \delta_1 + |\alpha|^2).
\end{array}
\end{equation}

First we note that in the limit $|y|=0$ the neutrino mass matrix
(\ref{mnu}) is diagonalized by a maximal rotation in the 23 plane, but
the ``solar sector'' is clearly inconsistent with experiment. On the
other hand, when $|\alpha|=0$ the unitary diagonalizing matrix can not
be fixed since there are two degenerate eigenvalues. Thus $|\alpha|\ne
0$ and $y\ne 0$ are required by experimental data.  In contrast $|x|$
can be taken equal to zero, in fact in this case the three eigenvalues
are distinct and the neutrino mass matrix $M_\nu$ is still
diagonalized by the $U_{\textrm{HPS}}$ matrix. The most relevant
phenomenological consequence of taking $|x|=0$ is that the
neutrinoless double beta decay amplitude $\propto m_{ee}$ is zero, see
Eq.\,(\ref{mee}).
%
%
 
\begin{figure}
\includegraphics[angle=0,width=90mm,height=6cm]{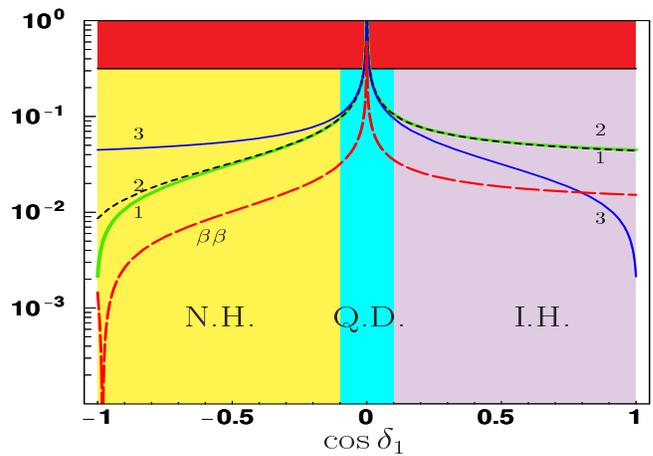}
\caption{$|m_{1}|$ (solid green), $|m_{2}|$ (small dashed dark),
  $|m_{3}|$ (solid blue) and $|m_{ee}|$ (large dashed red) vs $\cos
  \delta_1$ with $|\alpha|=1.1$ and $\cos \delta_2=1$. }\label{fig1}
\end{figure}

We can solve the system in Eq.\,(\ref{syst}) with respect to $|x|$ and
$|y|$ and write $m_{1,2,3}$ and $m_{ee}$ as function of
$\delta_1,\,\delta_2$ and $|\alpha|$.  For any value of
$\cos\delta_2$, we have three possibilities: {\it i)} for
$\cos\delta_1\sim 0$ we have $|m_1|\sim|m_3|$ that correspond to the
degenerate spectrum, {\it ii)} if $\cos\delta_1< 0$ we have
$|m_1|<|m_3|$ or normal hierarchy and {\it iii)} if $\cos\delta_1> 0$
we have $|m_1|>|m_3|$ or inverse hierarchy. These cases are
illustrated in Fig.\,(\ref{fig1}) where we plot the three masses
versus $\cos\delta_1$ for a fixed value of $|\alpha|$ and
$\cos\delta_2$.
We note that for $\alpha\sim 0$ we have only the degenerate case since
$|m_1|\sim |m_3|$.

The rate for neutrinoless double beta decay (\ref{mee})
vanishes when $|x|=0$. Solving the system in Eq.\,(\ref{syst})
for $|x|=0$ we find the relation
\begin{equation}\label{cos1}
\cos\delta_1=-\frac{3(1+|\alpha|^2)}{2|\alpha|(3+2 r)}
\end{equation}
where $r=\Delta m^2_{sol}/\Delta m^2_{atm}$. This special point
correspond to the narrow dip in Fig.\,(\ref{fig1}). By changing the
value of $|\alpha|$ in Eq.\,(\ref{cos1}) we have $-1\le\cos\delta_1<
-0.975$ for $0.8<|\alpha|<1.24$ (we have used the best fit value for
$r=0.032$, for different values the allowed range for $|\alpha|$ and
$\cos\delta_1$ changes only slightly). For other values of $|\alpha|$
and $\cos\delta_1$ the neutrinoless double beta decay rate can not
cancel.  All the general considerations given here are quite
independent of the value of $\cos\delta_2$ that is related to the
difference between $|m_{ee}|$ and $|m_1|$.  In fact if $\cos\delta_2
\sim 0$ then $|y|\ll |x|$ and $|m_{ee}|\sim|m_1|$. We have taken
$\cos\delta_2=1$ in Fig.\,(\ref{fig1}) since it gives the lowest value
of $m_{ee}$.

\vskip10mm We now turn to a theoretical derivation of the neutrino
mass matrix in Eq.\,(\ref{mnu}) based on the flavor symmetry $A_4$,
the finite group of the even permutations of four objects (for a short
introduction to $A_4$, see for instance \cite{Altarelli:2005yx} and
references therein). It has four irreducible representations, one
triplet $3$ and three singlets $1,1',1''$.  Let $\chi$ and $\varphi$
be two triplets of $A_4$ then
\begin{equation}\label{prod}
\begin{array}{c}
1\sim(\chi\varphi)  =(\chi _1\varphi_1+\chi_2\varphi_3+\chi_3\varphi_2),\\
1'\sim(\chi\varphi)' =(\chi_3\varphi_3+\chi_1\varphi_2+\chi_2\varphi_1),\\
1''\sim(\chi\varphi)''=(\chi_2\varphi_2+\chi_1\varphi_3+\chi_3\varphi_1),\\
3_s\sim
\left(
\begin{array}{c}
2\chi_1\varphi_1-\chi_2\varphi_3-\chi_3\varphi_2\\
2\chi_3\varphi_3-\chi_1\varphi_2-\chi_2\varphi_1\\
2\chi_2\varphi_2-\chi_1\varphi_3-\chi_3\varphi_1\\
\end{array}
\right),\,
3_a\sim
\left(
\begin{array}{c}
\chi_2\varphi_3-\chi_3\varphi_2\\
\chi_1\varphi_2-\chi_2\varphi_1\\
\chi_1\varphi_3-\chi_3\varphi_1
\end{array}
\right)
\end{array}
\end{equation}
and for instance \cite{FHLM}
$$1\sim(\chi\bar{\varphi})  =(\chi _1\varphi_1+\chi_2\varphi_2+\chi_3\varphi_3).$$

Consider as starting point the model defined in Table\,(\ref{tab1}).
We have three Higgs doublets $h_{1,2,3}$ in the $1$, $1''$ and $1'$
representations of $A_4$ and one Higgs triplet $\Delta_s$ transforming
as a singlet with respect to $A_4$.
\begin{table}[t]
\begin{tabular}{|c|cc|cc|}
\hline
&$L_i$&$l^c_i$&$h_{1,2,3}$&$\Delta_s$\\
\hline
$SU(2)$&$2$&$1$&$2$&$3$\\
$A_4$&$3$&$3$&$1,1'',1'$&$1$\\
\hline
\end{tabular}\caption{Lepton multiplet structure of our simple 
model, see text. }\label{tab1}
\end{table}
%
%
The $A_4$ invariant renormalizable Lagrangian reads
\begin{eqnarray}
\label{eq:LL}
\mathcal{L}&=&{y_{1_{ij}}}(L_il^c_j)h_{1}+y_{2_{ij}}(L_il^c_j)'h_{2}+y_{3_{ij}}(L_il^c_j)''h_{3}+\nonumber\\ 
&&+y_{ijk}^{\Delta_s}(L_iL_j)\Delta_{s_k}
\end{eqnarray}
where the structure of the Yukawa interaction can be easily obtained from
the product rule in Eq.\,(\ref{prod}).
Defining 
\begin{equation}\label{vev1}
\vev{ h_1}=v_1,\quad \vev{ h_2}=v_2,\quad \vev{ h_3}=v_3
\end{equation}
the charged lepton mass matrix is given as
\begin{equation}
M_l=\left(
\begin{array}{ccc}
a&b&c\\
b&c&a\\
c&a&b
\end{array}
\right)
\end{equation}
where $a=y_1 v_1$, $b=y_2 v_2$, $c=y_3 v_3$. It is diagonalized as
\begin{eqnarray}\label{Ml}
&U_\omega^\dagger M_l U_\omega^\dagger=
\left(
\begin{array}{ccc}
a+b+c&0&0\\
0& a+\omega b+\omega^2c&0\\
0&0&a+\omega^2b+\omega c
\end{array}
\right),&\nonumber\\
&U_\omega=\left(
\begin{array}{ccc}
1&1&1\\
1&\omega&\omega^2\\
1&\omega^2&\omega
\end{array}
\right)&
\end{eqnarray}
with $\omega^3=1$. As required, the charged lepton mass matrix has
three distinct eigenvalues.

From the Lagrangian in Eq.~(\ref{eq:LL}) we have the tree level
contribution to $M_\nu$ shown in Fig.\.(\ref{fig2}):
\begin{figure}[h!]
\includegraphics[angle=0,width=40mm]{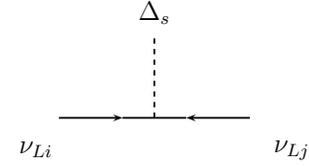}
\caption{Tree level contribution to $M_\nu$. }\label{fig2}
\end{figure}

This diagram is the type-II seesaw~\cite{schechter:1980gr}
contribution to $M_\nu$.  
The tree-level diagram yields
\begin{equation}\label{m0}
M_\nu^0=\left(
\begin{array}{ccc}
z&0&0\\
0&0&z\\
0&z&0
\end{array}
\right)
\end{equation}
where $z=y^{\Delta_s}v_s$.

\begin{figure}[t]
\includegraphics[angle=0,width=45mm]{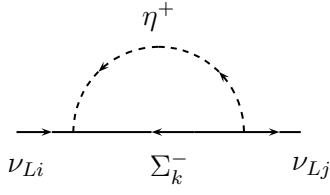}
\caption{One loop corrections to $M_\nu$ induced by fermion triplets $\Sigma_i$. }\label{fig33}
\end{figure}

Consider now adding an $A_4$ triplet of fermion isotriplets $\Sigma\sim 3$
with hypercharge zero, together with one Higgs doublet being a singlet 
under $A_4$, $\eta\sim 1$. Both $\Sigma$ and $\eta$ carry an extra $Z_2$ 
quantum number. Since $\Sigma$ is in the adjoint representation of the 
electroweak group, it has a Majorana mass term, see for instance \cite{Abada:2008ea}. 
We have the following extra terms in
the Lagrangian
\begin{equation}\label{LII}
y_{\Sigma_{ik}}\, (L_i  \Sigma_k) \eta + \mu_{kl}\,\Sigma_k \Sigma_l 
\end{equation}
where we assume $\mu$ to break softly $A_4$, that is 
\begin{equation}\label{mu1}
\mu=m_\Sigma \left(
\begin{array}{ccc}
1&0&0\\
0&\xi&1\\
0&1&\xi
\end{array}
\right).
\end{equation}
When $\xi\to 0$, $\mu$ respects $A_4$.
In general the fermion triplets $\Sigma_i$ induce 
one-loop neutrino wave-function corrections, see Fig.\.(\ref{fig33}),
giving
\begin{equation}\label{loopR}
R_{ij}=\frac{{y_\Sigma}^2}{16 \pi^2}\sum_kR^\mu_{ik}R^{\mu*}_{jk}B_1(m_{\Sigma_k}^2,m_\eta^2)
\end{equation}
where $m_{\Sigma_1}=m_{\Sigma }$, $m_{\Sigma_2}=m_{\Sigma} (-1+\xi)$, $m_{\Sigma_3}=m_{\Sigma} (1+\xi)$ 
are the eigenvalues of the matrix $\mu$ in Eq.\,(\ref{mu1}) and 
$$
R^\mu=\left(
\begin{array}{ccc}
1&0&0\\
0&1/\sqrt{2}&1/\sqrt{2}\\
0&-1/\sqrt{2} &1/\sqrt{2}
\end{array}
\right)
$$
is the orthogonal matrix that diagonalizes $\mu$. Since the fermion 
triplet masses are not degenerate, the loop correction is off-diagonal, 
\begin{equation}\label{Rb}
R=
\left(
\begin{array}{ccc}
\delta_1&0&0\\
0&\delta_2&\delta_3\\
0&\delta_3&\delta_2
\end{array}
\right)
\end{equation}
with
\begin{eqnarray}\label{Rp}
\delta_1&=&
\frac{{y_\Sigma}^2}{16 \pi^2}B_1(m_{\Sigma_1}^2,m_\eta^2),\nonumber\\
\delta_2&=&\frac{{y_\Sigma}^2}{32 \pi^2} 
                (B_1(m_{\Sigma_2}^2,m_\eta^2)+B_1(m_{\Sigma_3}^2,m_\eta^2)),
\nonumber\\
\delta_3&=&\frac{{y_\Sigma}^2}{32 \pi^2} 
(B_1(m_{\Sigma_3}^2,m_\eta^2)-B_1(m_{\Sigma_2}^2,m_\eta^2)),
\end{eqnarray}
The neutrino mass matrix $M_\nu^0$ in Eq.\,(\ref{m0}), is
corrected by the one loop diagram Fig.\,(\ref{fig33})  as follows
\begin{equation}\label{MR}
M_\nu=M_\nu^0 + R M_\nu^0 +M_\nu^0 R^T.
\end{equation}
After some algebra Eqs.\,(\ref{Ml}) and (\ref{MR}) are 
diagonalized by  $U_{\textrm{HPS}}$ and $M_\nu$ is of the form of Eq.\,(\ref{mnu}) 
as required.

\begin{figure}[t]
\includegraphics[angle=0,width=40mm]{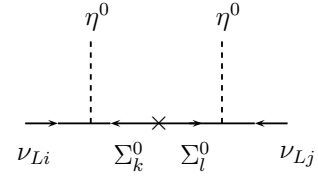}
\caption{Type III seesaw contribution to $M_\nu$. }\label{fig3}
\end{figure}

If the $Z_2$ symmetry is broken, $\eta$ takes a vacuum expectation 
value $\vev{\eta}=u$ and  the diagram of
Fig.\,(\ref{fig3}) 
gives a type-III
seesaw contribution \cite{Foot:1988aq} to the neutrino mass matrix where the heavy Majorana
neutrino of type-I seesaw is simply replaced with $\Sigma^0$ and is given by 
\begin{equation}\label{mu}
-y_\Sigma\cdot \mu^{-1} \cdot y_\Sigma^T.
\end{equation}
Then from Eqs.\,(\ref{m0}) and (\ref{mu}) we have
\begin{equation}\label{mnu2}
M_\nu=\left(
\begin{array}{ccc}
z+\beta&0&0\\
0&\frac{\beta\xi}{\xi^2-1}&z-\frac{\beta}{\xi^2-1}\\
0&  z-\frac{\beta}{\xi^2-1} &\frac{\beta\xi}{\xi^2-1}
\end{array}
\right)
\end{equation}
where $\beta=y_{\Sigma}^2 u^2/m_\Sigma$. Again $M_\nu$ is
diagonalized by the  $U_{\textrm{HPS}}$ in the basis where charged
leptons are diagonal since in such a basis it has the form of matrix in Eq.\,(\ref{mnu})
with the eigenvalues given by
\begin{equation}\label{mass}
\begin{array}{lll} 
m_1&=&\frac{\beta+z(1+\xi)}{1+\xi}\\
m_2&=&\beta+z\\
m_3&=&\frac{\beta+z(1-\xi)}{1-\xi}.
\end{array}
\end{equation}
Eqs.\,(\ref{mass}) are compatible with normal, degenerate and inverse neutrino mass spectra.
For instance when $\beta=-z(1+\xi)$ then $m_1=0$ and we have the simple relation
$$
4 r=|1-\xi|^2.
$$
If $\beta=-z(1-\xi)$ then $m_3=0$. This correspond to the inverse hierarchy and we have
$$
 r=1-\frac{4}{|1-\xi|^2}.
$$
In both cases we need $\xi\sim O(1)$ to reproduce data. For $|\xi|\ll 1$ we have  degenerate neutrino masses.

We mention in passing that in general one could also have the
situation where both tree-level, Fig.\,(\ref{fig3}), and wave function
correction, Fig.\,(\ref{fig33}), contribute to $M_\nu$, leading
however to the same overall structure in Eq.\,(\ref{mnu}). We
therefore do not discuss this ``mixed'' case in detail.

\vskip10mm
The above is just the simplest model with $U_{\textrm{HPS}}$ mixing that meets
all phenomenological requirements.
Other variants are easy to obtain. We could for example replace the
Higgs triplet with two Higgs doublets.  They contribute proportional
to $z$ to the neutrino mass matrix via a dimension five operator.

Finally we mention possible accelerator tests of our model
realization. If the members of the scalar triplet are light enough to
be produced at LHC \cite{Gunion:1996pq}, the decay pattern of the
$\Delta_s^{++}$ is completely fixed by the $A_4$ symmetry, 
and can be predicted to have only flavour diagonal decays
$Br(\Delta_s^{++}\to e^+e^+)=Br(\Delta_s^{++}\to
\mu^+\mu^+)=Br(\Delta_s^{++}\to \tau^+\tau^+)$ with all $Br(\Delta_s^{++}\to
l^+_i l^+_j)=0$.
This is different from, for example, the
case discussed in \cite{Chun:2003ej}, where the type-II seesaw is
assumed to be the only source of neutrino mass.  Also the cross
section for the fermionic triplet at LHC has been recently studied
\cite{Franceschini:2008pz,delAguila:2008cj}.  If $m_\Sigma \lsim 1\,TeV$,
$\Sigma^+\Sigma^-$ pairs will be produced decaying to $l^+l^-+2\eta$.
Again the branching ratios to the different lepton flavors are
completely fixed by the $A_4$ symmetry.  
Pair produced $\Sigma^{\pm}$ will decay to two charged leptons
plus two $\eta$ with branching ratios to the different lepton
flavours approximately fixed by our ansatz for A4 breaking.
Since the charged lepton mass matrix is diagonalized by $U_\omega$,
the final states can be predicted to be $Br(e\mu)\simeq Br(e\tau)$,
$Br(\mu\mu)\simeq Br(\tau\tau)$ and $Br(ee)\simeq Br(\mu\mu) +
Br(\mu\tau)-Br(e\tau)$.

%

\vskip10mm

In summary we have pointed out new ways to model $U_{\textrm{HPS}}$ lepton
mixing from first principles.  The minimal scheme is characterized by
only three parameters and requires $m_{ee}=0$.  Five parameter schemes
can produce an arbitrary neutrino mass spectrum where $m_{ee}$ is not
strictly predicted but related to the two Majorana phases of the
lepton mixing matrix. We have discussed one model realization of these
schemes, which interestingly avoids the vev misalignment problem
present in other models based on $A_4$, as a consequence of not using  
$A_4$ triplet Higgses. This model is potentially
testable at the LHC if either the scalar triplet or the fermionic
triplet (or both) have mass below approximately $1\,TeV$.

\vskip .2cm {\bf Acknowledgments} \vskip .2cm 
We thank L. Lavoura and F. Bazzocchi for useful comments and discussions. 
Work supported by MEC grant FPA2008-00319/FPA, by European Commission Contracts
MRTN-CT-2004-503369 and ILIAS/N6 RII3-CT-2004-506222.

 \def\baselinestretch{1}


\end{document}